\documentclass[aps,prx,twocolumn,reprint,superscriptaddress]{revtex4-1}
\usepackage{amsmath}
\usepackage{amssymb}
\usepackage{graphicx}
\usepackage{hyperref}
\usepackage{multirow}
\usepackage{color}

\begin{document}

\title{Cycloidal magnetism driven ferroelectricity in double tungstate LiFe(WO$_4$)$_2$}
\author{Meifeng Liu}
\affiliation{Institute for Advanced Materials and School of Physics and Electronic Science, Hubei Normal University, Huangshi 435002, China}
\affiliation{Laboratory of Solid State Microstructures and Innovative Center of Advanced Microstructures, Nanjing University, Nanjing 210093, China}
\author{Lingfang Lin}
\author{Yang Zhang}
\affiliation{Department of Physics, Southeast University, Nanjing 211189, China}
\author{Shaozhen Li}
\affiliation{School of Physics and Institute for Quantum Materials, Hubei Polytechnic University, Huangshi 435003, China}
\author{Qingzhen Huang}
\affiliation{NIST Center for Neutron Research, National Institute of Standards and Technology, Gaithersburg, Maryland 20899, USA}
\author{V. Ovidiu Garlea}
\author{Tao Zou}
\affiliation{Quantum Condensed Matter Division, Oak Ridge National Laboratory, Oak Ridge, TN 37831, USA}
\author{Yunlong Xie}
\affiliation{Laboratory of Solid State Microstructures and Innovative Center of Advanced Microstructures, Nanjing University, Nanjing 210093, China}
\author{Yu Wang}
\affiliation{Institute for Advanced Materials and School of Physics and Electronic Science, Hubei Normal University, Huangshi 435002, China}
\author{Chengliang Lu}
\affiliation{School of Physics, Huazhong University of Science and Technology,Wuhan 430074, China.}
\author{Lin Yang}
\affiliation{Institute for Advanced Materials and Laboratory of Quantum Engineering and Materials, South China Normal University, Guangzhou 510006, China}
\author{Zhibo Yan}
\affiliation{Laboratory of Solid State Microstructures and Innovative Center of Advanced Microstructures, Nanjing University, Nanjing 210093, China}
\author{Xiuzhang Wang}
\affiliation{Institute for Advanced Materials and School of Physics and Electronic Science, Hubei Normal University, Huangshi 435002, China}
\author{Shuai Dong}
\email{sdong@seu.edu.cn}
\affiliation{Department of Physics, Southeast University, Nanjing 211189, China}
\author{Jun-Ming Liu}
\affiliation{Institute for Advanced Materials and School of Physics and Electronic Science, Hubei Normal University, Huangshi 435002, China}
\affiliation{Laboratory of Solid State Microstructures and Innovative Center of Advanced Microstructures, Nanjing University, Nanjing 210093, China}
\affiliation{Institute for Advanced Materials and Laboratory of Quantum Engineering and Materials, South China Normal University, Guangzhou 510006, China}
\date{\today}

\begin{abstract}
Tungstates $A$WO$_4$ with the wolframite structure characterized by the $A$O$_6$ octahedral zigzag chains along the $c$-axis, can be magnetic if $A$=Mn, Fe, Co, Cu, Ni. Among them, MnWO$_4$ is a unique member with a cycloid Mn$^{2+}$ spin order developed at low temperature, leading to an interesting type-II multiferroic behavior. However, so far no other multiferroic material in the tungstate family has been found. In this work, we present the synthesis and the systematic study of the double tungstate LiFe(WO$_4$)$_2$. Experimental characterizations including structural, thermodynamic, magnetic, neutron powder diffraction, and pyroelectric measurements, unambiguously confirm that LiFe(WO$_4$)$_2$ is the secondly found multiferroic system in the tungstate family. The cycloidal magnetism driven ferroelectricity is also verified by density functional theory calculations. Although here the magnetic couplings between Fe ions are indirect, namely via the so-called super-super-exchanges, the temperatures of magnetic and ferroelectric transitions are surprisingly much higher than those of MnWO$_4$.
\end{abstract}
\maketitle

\section{Introduction}
Frustrated magnetism plays a key role in the so-called type-II multiferroics \cite{Khomskii:Phy}, in which ferroelectric polarizations ($P$) are triggered by particular magnetic orders \cite{Cheong:Nm,Wang:Ap,Dong:Ap}. Not only some noncollinear spin orders (e.g. the cycloidal type) but also some collinear spin orders (e.g. the up-up-down-down type) can break the spatial inversion symmetry. The corresponding microscopic magnetoelectric mechanisms are the inverse Dzyaloshinskii-Moriya (DM) interaction and exchange striction, respectively \cite{Sergienko:Prb,Katsura:Prl,Mostovoy:Prl06,Sergienko:Prl,Choi:Prl}. Despite the diversification of routes to magnetoelectricity, these frustrated magnetic orders can compete and even coexist in some systems. For example, in the most studied orthorhombic $R$MnO$_3$, the ground state changes from the cycloidal antiferromagnetism (AFM) (e.g. $R$=Tb, Dy) to the collinear E-type antiferromagnetism (e.g. $R$=Ho, Y) \cite{Goto:Prl,Dong:Prb08.2,Dong:Mplb}. Even in those canonical cycloidal systems, e.g. DyMnO$_3$, the synchronization of Dy's and Mn's magnetic moments lead to the exchange striction effect \cite{Zhang:Apl11,Zhang:Apl11.2}. These multiple magnetoelectric orders also exist in other complex Mn-oxides, e.g. $R$Mn$_2$O$_5$ \cite{Hur:Nat,Chapon:Prl,Lee:Prl13,Zhao:Sr} and CaMn$_7$O$_{12}$ \cite{Zhang:Prb11,Johnson:Prl,Lu:Prl}, which lead to plethoric multiferroic physics as well as better magnetoelectric performances due to combined benefits. 

Besides these well-studied Mn-oxides, the tungstate family with the wolframite structure is another playground with plenty frustrated magnetic orders. For example, MnWO$_4$ is a multiferroic material when temperature ($T$) is in the range $7.6$ K to $12.7$ K, corresponding to the incommensurate elliptical spiral phase. Below this $T$ range the system displays a commensurate collinear antiferromagnetic state \cite{Taniguchi:Prl,Heyer:Jpcm}. The magnetic field and ion substitutions can significantly tune the magnetism and thus the associated ferroelectricity \cite{Taniguchi:Prl09,Chaudhury:Prb,Song:Prb09,Yu:Prb13}. It is worth noting that the other tungstate members, e.g. FeWO$_4$, CoWO$_4$, NiWO$_4$, CuWO$_4$, and NaCr(WO$_4$)$_2$, display collinear antiferromagnetic orders and are not multiferroics \cite{Ehrenberg:jpcm,Forsyth:jpcm,Wilkinson:Kristallogr,Dey:IC}.

Very recently, Holbein \textit{et al.} reported a double tungstate NaFe(WO$_4$)$_2$ to exhibit a three-dimensional incommensurate spiral spin structure below $4$ K, which could be tuned to a commensurate collinear spin structure by applying magnetic field \cite{Holbein:Prb}. However, their neutron study revealed the condensation of a single irreducible representation of magnetic structure, which did not imply a nonzero ferroelectric $P$ because spirals with opposite chirality coexisted. For comparison, MnWO$_4$ exhibits a magnetic structure derived from two irreducible representations, which leads a finite ferroelectric polarization ($P$). Nevertheless, this work is deserved of attention because it raises a possibility of searching new multiferroics from the $AA'$(WO$_4$)$_2$ sub-family (i.e. double tungstate) where $A$ or $A'$ is magnetic.

\begin{figure}
\centering
\includegraphics[width=0.4\textwidth]{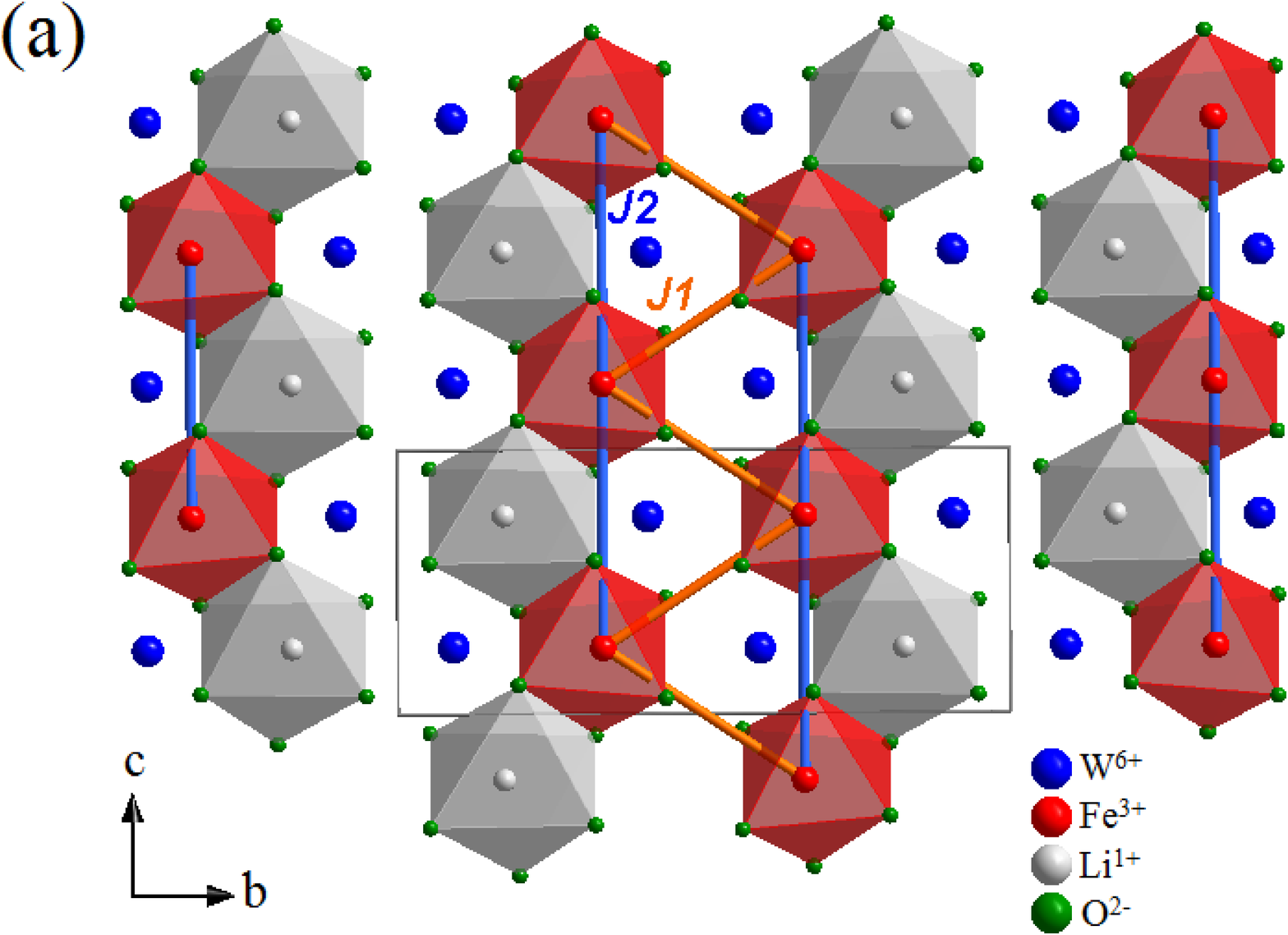}
\includegraphics[width=0.4\textwidth]{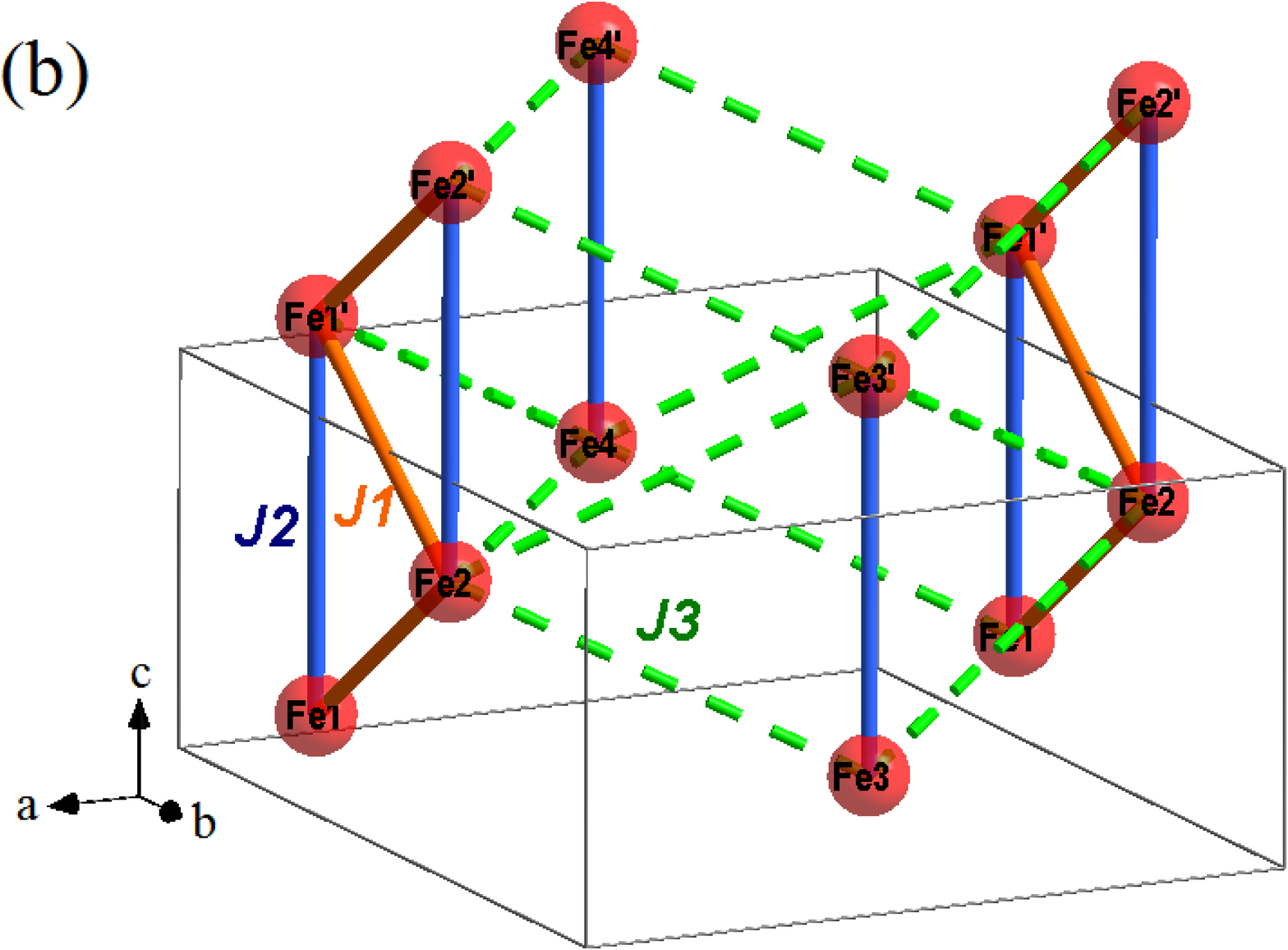}
\includegraphics[width=0.45\textwidth]{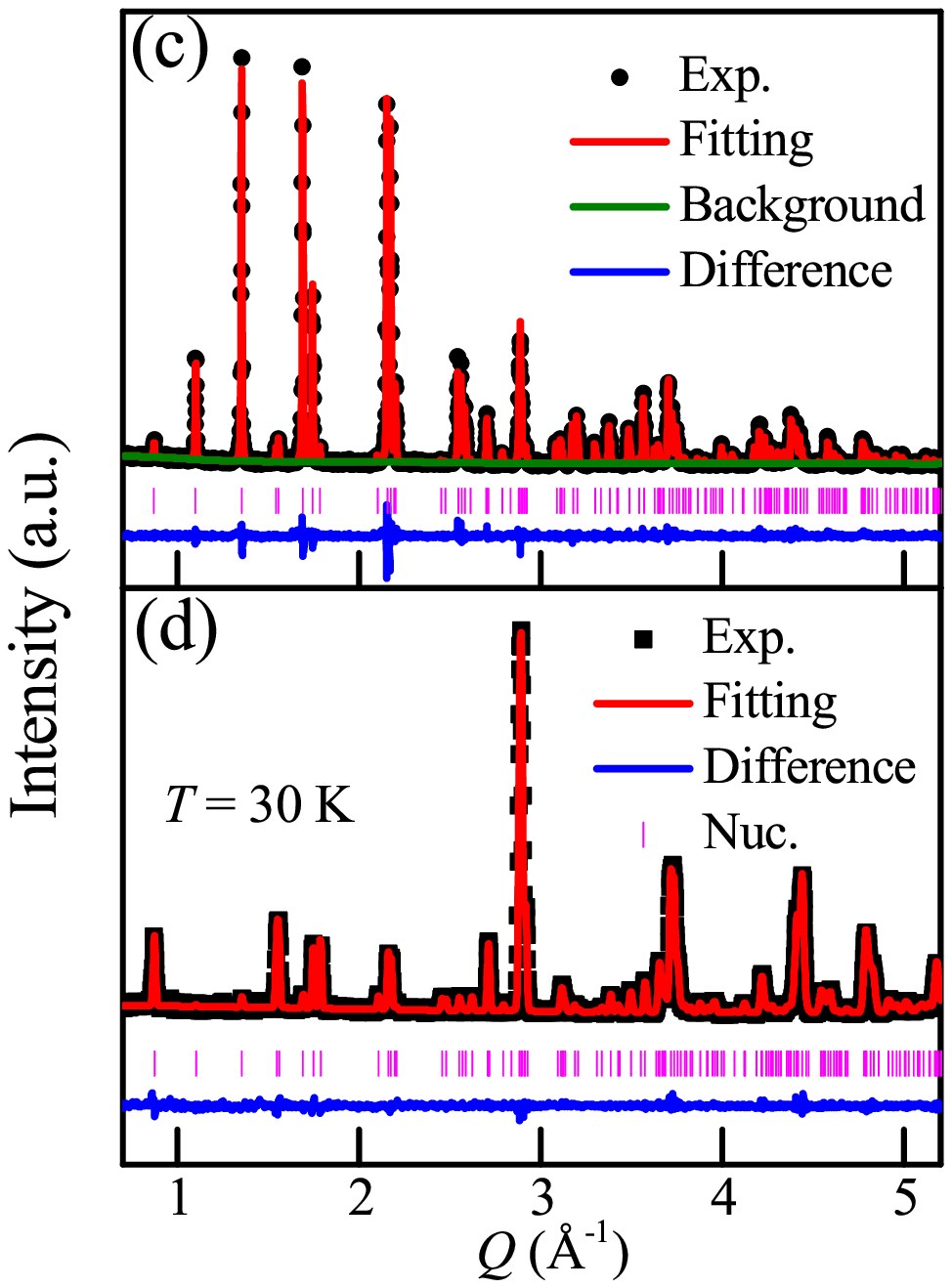}
\caption{(a) Projection in the $bc$ plane of the crystal structure of LiFe(WO$_4$)$_2$. Blue: W; green: O; grey: Li; red: Fe. Noting the arrangement of Li/Fe/W is different from those in NaFe(WO$_4$)$_2$ \cite{Holbein:Prb}. (b) The framework of Fe ions and the magnetic exchange paths $J_1$/$J_2$/$J_3$. (c) The XRD pattern measured at room $T$ and the corresponding Rietveld fit. (d) Rietveld fit of the NPD pattern measured at $30$ K.}
\label{Fig1}
\end{figure}

Here we reported another double tungstate LiFe(WO$_4$)$_2$ which is very similar to NaFe(WO$_4$)$_2$ in chemical composition and stoichiometry. However, the fine difference between the crystallographic structure of these two materials leads to different magnetoelectric result. Our systematical experimental characterizations and theoretical calculations confirm that LiFe(WO$_4$)$_2$ is a new multiferroic system, the second multiferroic member in the tungstate family. More surprisingly, the magnetic and ferroelectric transitions in LiFe(WO$_4$)$_2$ is even much higher than those in MnWO$_4$ \cite{Taniguchi:Prl,Heyer:Jpcm}, although the indirect exchange paths Fe-O-O-Fe (or Fe-O-W-O-Fe) seem to more complicated comparing with the Mn-O-Mn paths in MnWO$_4$.

\section{Methods}
\subsection{Details of experimental process}
Polycrystalline samples of LiFe(WO$_4$)$_2$ were prepared by the convention solid sintering method, using the highly purified powder of oxides and carbonates as starting materials. The stoichiometric mixtures were ground and fired at $600$ $^{\circ}$C for $24$ hours in air. The resultant powder was re-ground and pelletized under a pressure of $1000$ psi into disks of $2.0$ cm in diameter, and then these pellets were sintered at $750$ $^{\circ}$C for $24$ hours in air in prior to natural cooling down to room $T$. Phase purity of samples were checked by X-ray diffraction (XRD) at room $T$ using the Cu-$K$ radiation of X-ray power diffractometer (D8 advanced, Bruker).

The crystalline and magnetic structures are also checked using neutron powder diffraction (NPD) carried out at NIST using the BT1 powder diffractometer. The NPD patterns were collected with neutron wavelength $\lambda$ = $2.0775$ {\AA} at $T$ = $5$ K and $30$ K. The NPD data were analyzed using the Rietveld refinement program FULLPROF \cite{Rodriguez:Pb}.

The dc magnetization as a function of $T$ and magnetic field ($H$) were measured using the Quantum Design superconducting quantum interference device magnetometer (SQUID). The specific heat was measured in the $T$ range of $2$ K to $300$ K and under magnetic fields up to $9$ T by the physical property measurement system (PPMS, Quantum Design) using the relaxation method.

For the electrical measurements, the disk-like samples of $3.0$ mm in diameter and $0.2$ mm in thickness were deposited with Au electrodes on the top/bottom surfaces. The dielectric constant as a function of $T$ was measured using the HP4294A impedance analyzer attached to the PPMS. In our experiments, both Au and silver paste electrodes were used and no observable difference was found regarding the dielectric constant and pyroelectric current. The ferroelectric $P$ was measured by integrating zero-electric-field pyroelectric current ($I_{pyro}$) recorded using the Keithley $6514$ electrometer. For the pyroelectric current measurement, the sample was electrically poled under an electric field $E_{\rm p}=10$ kV/cm and cooled down from $50$ K to $2$ K. Then the poling field was removed and the sample was electrically short-circuited for a long time (e.g. several hours) in order to exclude possible extrinsic contributions (e.g. trapped charge). The background of $I_{pyro}$ was reduced to less than $0.2$ pA. Different warming rates, e.g. $2-6$ K/min, was adopted to record $I_{pyro}$, which gave an identical peak position and value of $P$ in tolerable precision.

\subsection{Details of DFT calculations}
Density functional theory (DFT) calculations are performed using the projector augmented wave (PAW) pseudopotentials as implemented in Vienna {\it ab initio} Simulation Package (VASP) code \cite{Kresse:Prb,Kresse:Prb96,Kresse:Prb99,Perdew:Prl}. To acquire more accurate description of crystalline structure and electron correlation, the revised Perdew-Burke-Ernzerhof for solids (PBEsol) function and the generalized gradient approximation plus $U$ (GGA+$U$) method are adopted \cite{Perdew:Prl,Perdew:Prl08}. According to literature \cite{Zhang:Prb15,Weng:Prl}, the on-site Coulomb $U_{\rm eff}=U-J=4$ eV is applied to the $3d$ orbital of Fe, using the Dudarev implemention \cite{Dudarev:Prb}. The cutoff of plane wave basis was fixed to $625$ eV, a quite high value due to Li. The Monkhorst-Pack $k$-point mesh is $3\times$3$\times7$ for the minimal cell and correspondingly reduced for supercell calculations. To account the noncollinear spin texture, the spin-orbit coupling (SOC) is also switched on in calculations and the standard Berry phase method is adopted to estimate the ferroelectric $P$ \cite{Resta:Rmp}.

\section{Results \& discussion}
\subsection{Crystalline structure}
The crystal structure of LiFe(WO$_4$)$_2$ is described in the monoclinic space group $C2/c$ and consists of stacking (100) layers made of mixed [LiO$_6$] and [FeO$_6$] edge-sharing octahedra arranged in zigzag chains, separated by layers composed of tungstate [WO$_6$] octahedra. As noted in Ref.~\cite{Spinnler:JACeS}, the ordering in the wolframite-type phases in respect to the Li-Fe-W and Na-Fe-W arrangements is not the same. In NaFe(WO$_4$)$_2$ (described by $P2/c$ symmetry) each populated octahedral chain contains only one type of cation, while in LiFe(WO$_4$)$_2$ the chain contain both Li and Fe octahedra alternating along the $c$ direction. Such atomic arrangement leads to the doubling of the unit cell along the $b$-direction, as compared to the NaFe(WO$_4$)$_2$.

Figures~\ref{Fig1}(a-b) show the crystal structure of LiFe(WO$_4$)$_2$ as well as the magnetic framework of Fe$^{3+}$ ions. As visible in Fig.~\ref{Fig1}(a), the magnetic framework in LiFe(WO$_4$)$_2$ is no longer defined by zigzag chains as in the case of NaFe(WO$_4$)$_2$ and Mn(WO$_4$), but rather by frustrated two-leg spin ladders. Figure~\ref{Fig1}(c) shows the room $T$ powder XRD pattern of LiFe(WO$_4$)$_2$. A Rietveld refinement has been performed using the structural model proposed by Klevtsov~\cite{Klevtsov:icsd}, which provided satisfactory residual values ($R_p=8.51\%$, $R_{wp}=6.69\%$, and ${\chi}^2=1.952$). There is no impurity phase detected in the XRD power pattern. The crystalline structure is monoclinic with $a=9.2997$ {\AA}, $b=11.4302$ {\AA}, $c=4.9072$ {\AA}, $\beta=90.65^\circ$, in agreement with previous works \cite{Spinnler:JACeS, Klevtsov:icsd}. Figure~\ref{Fig1}(d) shows the Rietveld fit of the NPD pattern measured at $30$ K. The refined structural parameters of LiFe(WO$_4$)$_2$, such as lattice parameters, atomic coordinates and displacement parameters, resulted from the NPD data are listed in Table~\ref{Table1}.

\begin{table}
\caption{\label{Table1} Refined structural parameters of LiFe(WO$_4$)$_2$ from powder neutron diffraction data collected at $30$ K.}
\begin{ruledtabular}
\begin{tabular}{ccccc}
\centering{Atom (Wyck.)} & $x$ & $y$ & $z$ & B \\[3pt]
\hline
\centering {W~~($8f$)}& 0.2468(2) & 0.0910(4) & 0.250(1) & 0.34(7) \\[3pt]
\centering {Fe~~($4e$)}& 0 & 0.3359(3) & 0.25 & 0.1(1) \\[3pt]
\centering {Li~~($4e$)}& 0.5 & 0.342(1) & 0.25 & 0.7(2) \\[3pt]
\centering {O1~~($8f$)}&  0.3634(4)  & 0.0586(3)&  0.9225(6)&  0.12(2)\\[3pt]
\centering {O2~~($8f$)}& 0.3806(3) & 0.1821(3) & 0.4127(6) & 0.12(2) ) \\[3pt]
\centering {O3~~($8f$)}& 0.3559(3) & 0.5483(3) & 0.9451(6)&  0.12(2) \\[3pt]
\centering {O4~~($8f$)}&  0.3773(4) & 0.6946(2) & 0.3936(6) & 0.12(2) \\[3pt]
\hline \\
\multicolumn{5}{c}{SP: $C2/c$,~ $a$ = 9.2687(1)~\AA,~$b$ = 11.3964(1)~\AA,}\\[3pt]
\multicolumn{5}{c}{~$c$ = 4.89368(6)~\AA,~$\beta$=90.564(1)~deg.,~Chi$^2$ = 1.1}\\[3pt]
\multicolumn{5}{c}{R$_{Bragg}$ = 4.3$\%$,~R$_{p}$= 6.26$\%$,~$R_{wp}$= 7.84$\%$}\\[3pt]
\end{tabular}
\end{ruledtabular}
\end{table}

\subsection{Magnetic behavior}

\begin{figure}
\centering
\includegraphics[width=0.48\textwidth]{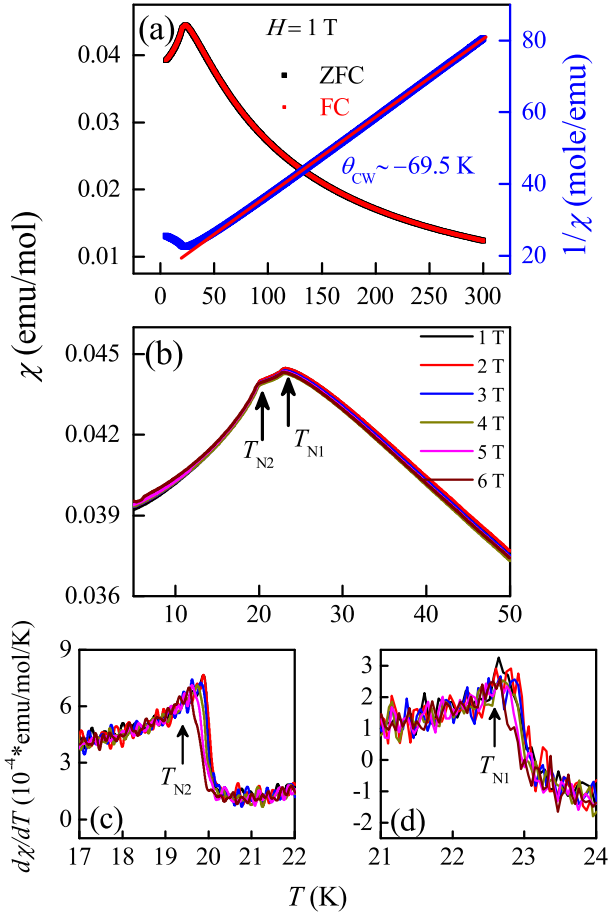}
\caption{Magnetism of LiFe(WO$_4$)$_2$. (a) Magnetic susceptibility $\chi$ (left) and Curie-Weiss fitting of $1/\chi$ (right). Both the FC and ZFC curves are shown, which are identically coinciding. (b) The amplified view around the phase transitions. (c-d) The amplified view of the derivatives of $\chi$ around the phase transitions.}
\label{Fig2}
\end{figure}

Figure~\ref{Fig2}(a) shows the $T$-dependent magnetic susceptibility $\chi$($T$) measured with a small field $H\sim0.1$ T over the $T$-range from $5$ K to $300$ K. The zero-field cooling (ZFC) and field cooling (FC) curves of $\chi$($T$) overlap in the whole measuring $T$ range. The fit of ${\chi}^{-1}$($T$) using the linear Curie-Weiss law yields a negative Curie-Weiss $\theta_{\rm CW}=-69.5$ K, suggesting strong antiferromagnetic interactions between Fe$^{3+}$ spins. An effective paramagnetic moment of $6.075$ $\mu_{\rm B}$ per Fe is found, which is very close to the expected value of spin-only effective moment ($5.92$ $\mu_{\rm B}$) for high-spin Fe$^{3+}$ ($S^z=5/2$, $L$ = 0).

Two successive magnetic transitions are observed at $T_{\rm N1}=22.6$ K and $T_{\rm N2}=19.7$ K, as evidenced by peaks of $\chi$($T$) and $d\chi$/$dT$, shown in Figs.~\ref{Fig2}(b-d). The ratio $\theta_{\rm CW}$/$T_{\rm N1}$ is $3.08$, indicating a moderate level of magnetic frustration. In addition, $\chi$($T$) under different $H$ are measured, as plotted in  Fig.~\ref{Fig2}(b). The large fields are only slightly affecting the values of $T_{\rm N1}$ and $T_{\rm N2}$.

These low-temperature antiferromagnetic transitions are reasonable considering the nearly-isolated [FeO$_6$]'s which are separated by [WO$_6$]'s and [LiO$_6$]'s. The exchanges between Fe spins can only be mediated via oxygen and tungsten, i.e. the so-called super-superexchanges Fe-O-O-Fe and Fe-O-W-O-Fe. Such complicated exchange routes suppress the effective strength of magnetic couplings. Even though, surprisingly, these antiferromagnetic transitions are still much higher than those in MnWO$_4$. Empirically, in general, the superexchange between Fe$^{3+}$ pair is much stronger than that between Mn$^{2+}$ pair. Ferrites, e.g. BiFeO$_3$, often show much higher magnetic transitions than Mn-based oxides.

\begin{figure}
\centering
\includegraphics[width=0.42\textwidth]{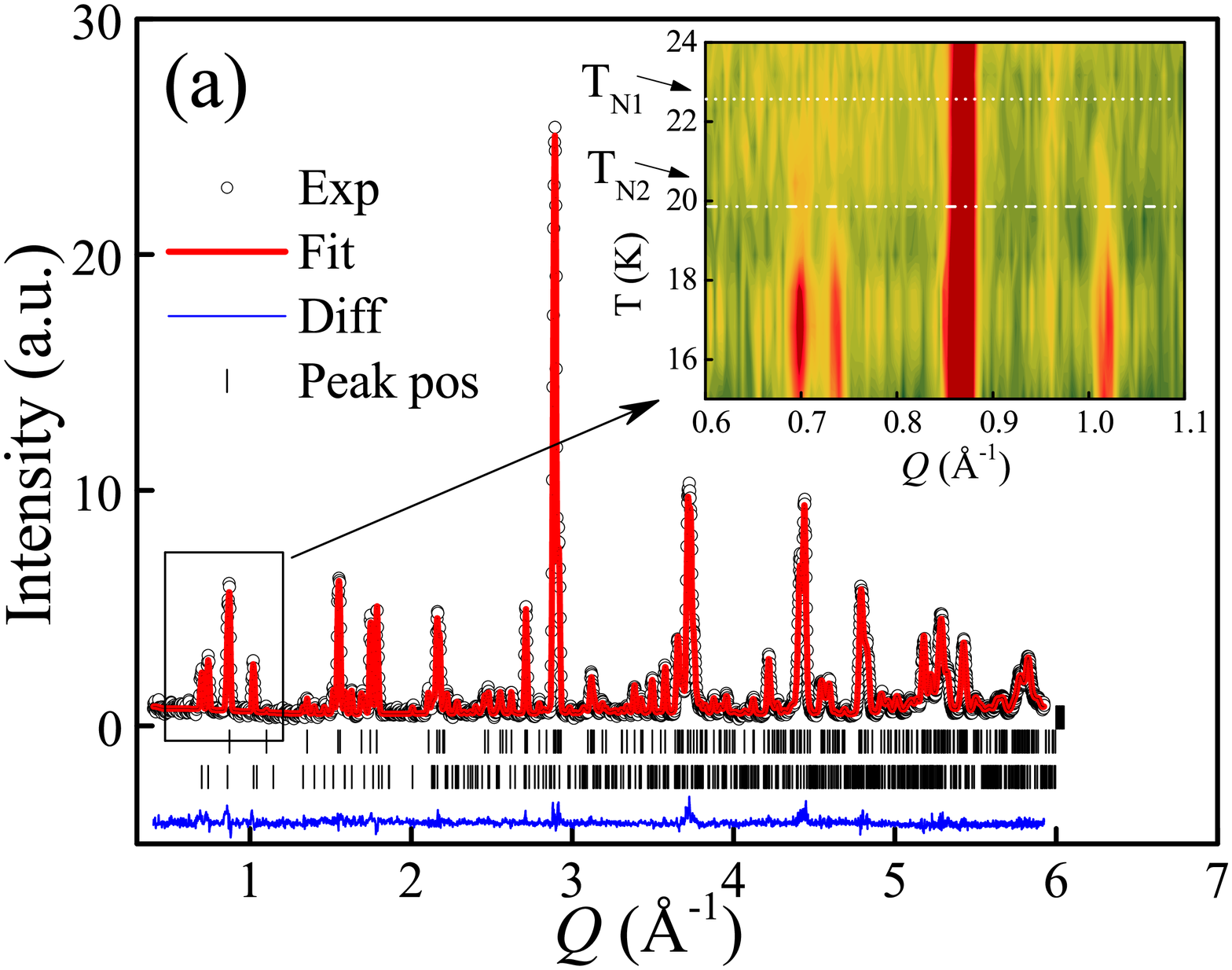}
\includegraphics[width=0.48\textwidth]{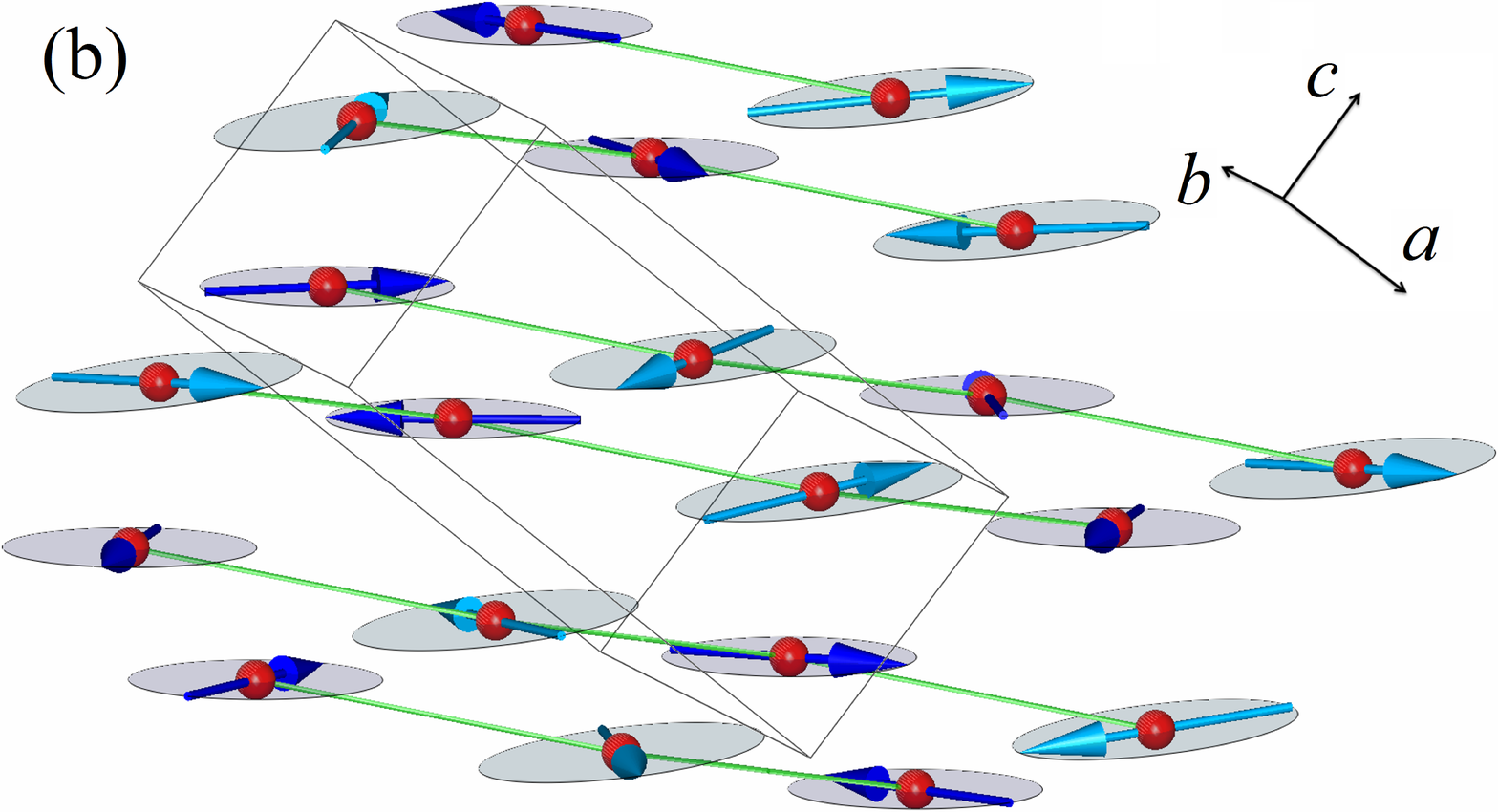}
\caption{(a) The NPD pattern measured at $5$ K and corresponding Rietveld fit. Inset: contour plot of the temperature dependence of magnetic Bragg peaks at small momentum transfer $\textbf{Q}$. (b) The sketch of noncollinear magnetic order fitted from the NPD data. The moments of Fe form a cycloidal structure with iron magnetic moments nearly confined to the plane defined by the $\textbf{k}$ vector and [010] direction. The cycloid rolls along the $J_3$ magnetic path depicted by solid green lines. . The moments at the two Fe positions related by the twofold axis symmetry, Fe1 ($0$, $y$, $1/4$), Fe2 ($0$, $-y$, $3/4$), are depicted by different colors.}
\label{Fig3}
\end{figure}

NPD is a powerful tool to reveal the frustrated magnetism in multiferroics \cite{Ratcliff:qm}. To reveal the magnetic ground state of LiFe(WO$_4$)$_2$, the NPD measurements were performed below $30$ K, as shown in Fig.~\ref{Fig3}(a). Comparing with the NPD pattern at $30$ K, no obvious change of lattice parameters has been observed but additional sharp magnetic Bragg reflections appear below approximately $20$ K, reflecting the presence of long range magnetic ordering. A contour plot showing the $T$ dependence of the magnetic scattering at low momentum transfer ($\textbf{Q}$) is given in the inset of Fig.~\ref{Fig3}(a). Neutron diffraction results suggest that the first transition observed in the macroscopic measurements at $T_{\rm N1}=22.6$ K corresponds to a short-range magnetic ordering, and that the long range order only forms below the second transition point $T_{\rm N2}=19.7$ K. The magnetic peaks that are present at the lowest measured $T$, $5$ K, can be successively indexed with an incommensurate propagation vector $\textbf{k}=$($0.890$, $0$, $0.332$). Interestingly to point out is that the other double tungstate NaFe(WO$_4$)$_2$ was also found to order with an incommensurate wave-vector $\textbf{k}=$($0.485$, $0.5$, $0.48$) \textit{et al.}. Furthermore, the doubling of the magnetic lattice along the $b$-direction in NaFe(WO$_4$)$_2$ is perfectly compatible with our wave-vector solution, considering that the size of the crystallographic unit cell in LiFe(WO$_4$)$_2$ is already doubled.

To determine the symmetry-allowed magnetic structures, given the crystal structure and the aforementioned propagation vector, representational analysis has been performed using $SARAh$-Representational Analysis. There are two possible irreducible representation (IRs) allowed for the Fe ion at the $4e$ Wyckoff position, corresponding to $\Gamma1$ and $\Gamma2$ in the Kovalev numbering scheme, as summarized in Table~\ref{TableII}. We found that two IRs are required to describe the incommensurate magnetic structure at $5$ K. The best solution was achieved when combining $\psi_2$ of $\Gamma1$ and $\psi_4$, $\psi_6$ of $\Gamma2$. The simultaneous appearance of both $\Gamma1$ and $\Gamma2$ are quite nontrivial, and it distinguishes the magnetism of LiFe(WO$_4$)$_2$ from the recently studied NaFe(WO$_4$)$_2$, although their chemical components are very similar \cite{Holbein:Prb}. We reiterate that no structural distortion has been detected to accompany this magnetic transition. Nevertheless, taking into account the succession of two second-order transitions, the magnetic ground state defined by two IRs does not violate the Landau theory of second order phase transitions. The refined magnetic structure of LiFe(WO$_4$)$_2$ is shown in Fig.~\ref{Fig3}(b). The magnetic moments of Fe form a cycloidal magnetic structure with the spins confined to the plane defined by the $\textbf{k}$ vector and [010] direction. The envelope of the cycloid is nearly circular, with a refined amplitude of the magnetic moment of $4.2(1)$ $\mu_{\rm B}$. This value is smaller than the expected value for the spin $S=5/2$, but it is not unreasonable considering the presence of frustrated magnetic interactions.

We would like to point out that the NPD is occasionally recognized as not sufficient to distinguish between the cycloid and spiral magnetic structures and therefore the single crystals measurements become highly desirable \cite{Johnson:Armr}. It all depends on number of non-equivalent magnetic sites and number of degrees of freedom in the model. Here we tested all possible symmetry-constrained models, which included cycloidal and spiral configurations, and found very different outcomes. The magnetic structure model described in the manuscript gives the best fitting to the NPD data.

\begin{table}[tbp]
\centering
\caption{Basis vectors for the space group $C12/c1$ with ${\bf k}$=($0.89$, $0$, $0.332$). The decomposition of the magnetic representation for the Fe site ($0$, $0.335$, $0.25$) is $\Gamma_{\rm Mag}=3\Gamma_1^1+3\Gamma_2^1$. The atoms of the nonprimitive basis are defined according to 1: ($0$, $0.335$, $0.25$), 2: ($0$, $0.665$, $0.75$). The $\zeta$ parameter is $\exp(-2\pi i{\bf k\cdot t})$.}
\label{TableII}
\begin{ruledtabular}
\begin{tabular}{cccc}
IR            & BV          & Fe1 (0,~0.335,~0.25) & Fe2 (0,~0.665,~0.75) \\
\hline
$\Gamma_{1}$ & $~\psi_{1}$ &  (1,~0,~0)    & (-$\zeta$,~0,~0)  \\
            & $~\psi_{2}$ &  (0,~1,~0)    & (0,~$\zeta$,~0)  \\
            & $~\psi_{3}$ &  (0,~0,~1)    & (0,~0,~-$\zeta$)  \\ \hline
$\Gamma_{2}$ & $~\psi_{4}$ &  (1,~0,~0)    & ($\zeta$,~0,~0)  \\
            & $~\psi_{5}$ &  (0,~1,~0)    & (0,~-$\zeta$,~0)  \\
            & $~\psi_{6}$ &  (0,~0,~1)    & (0,~0,~$\zeta$)  \\
\end{tabular}
\end{ruledtabular}
\end{table}

\subsection{Specific heat}

\begin{figure}
\centering
\includegraphics[width=0.48\textwidth]{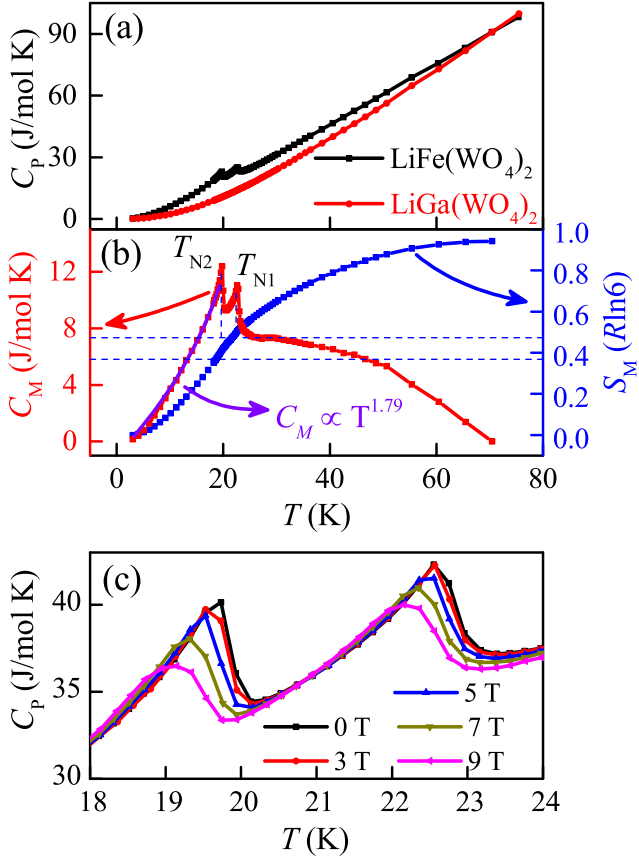}
\caption{(a) Heat capacities $C_{\rm P's}$ of magnetic LiFe(WO$_4$)$_2$ and nonmagnetic LiGa(WO$_4$)$_2$. (b) Left: the pure magnetic contribution to heat capacity $C_{\rm M}$, calculated by deducting the capacity of  LiGa(WO$_4$)$_2$ from that of LiFe(WO$_4$)$_2$. Right: the magnetic entropy. (c) Heat capacities measured under magnetic fields.}
\label{Fig4}
\end{figure}

To further characterize the phase transitions, the specific heat ($C_{\rm P}$) of LiFe(WO$_4$)$_2$ were measured. Figure~\ref{Fig4}(a) shows the curve of $C_{\rm P}$ in zero magnetic field as a function of $T$. Two pronounced peaks at $T_{\rm N1}=22.6$ K and $T_{\rm N2}=19.7$ K was observed, confirming the results of above magnetization measurements. Since LiFeW$_2$O$_8$ is a magnetic insulator, the specific heat contains the contributions from both magnons and phonons. In order to deduct the phonon contribution, the specific heat of the isostructural but nonmagnetic compound LiGa(WO$_4$)$_2$ was measured for reference, as shown in Fig.~\ref{Fig4}(a) too. Then the magnetic contribution $C_{\rm M}$($T$) can be estimated by subtracting the $C_{\rm P}$($T$) data of LiGa(WO$_4$)$_2$ from that of LiFe(WO$_4$)$_2$, as shown in Fig.~\ref{Fig4}(b). The magnetic entropy $S_{\rm M}$($T$) obtained by integrating the $C_{\rm M}$/$T$ data is also shown in Fig.~\ref{Fig4}(b), giving a saturation value of  $S_{\rm M}=0.94Rln6=14.003$ J/mol$\cdot$K at $70$ K, very close to the standard estimation $nRln(2S+1)=14.896$ J/mol$\cdot$K, where $n$ is the number of magnetic ion in one unit cell (here $n=1$).

Interestingly, the corresponding magnetic entropy gain at $T_{\rm N1}$ is only about $49\%$ of its saturated value, implying the release of partial magnetic entropy above $T_{\rm N1}$. Such a behavior suggests the persistence of short-range magnetic order even above $T_{\rm N1}$. In addition, the magnetic specific heat $C_{\rm Mag}$ below $T_{\rm N1}$ can be well fitted by the power law $C_{\rm Mag}$ = $AT^{\rm \alpha}$, where ${\rm \alpha}$ ${\sim}1.79$. This power law indicates that the magnetic lattice of LiFe(WO$_4$)$_2$ is pseudo one dimensional.

The field dependent specific heat $C_{\rm P}$($T$) is shown in Fig.~\ref{Fig4}(c). With increasing magnetic field the $T_{\rm N1}$ and $T_{\rm N2}$ transition points shift to lower values and the peaks in the $C_{\rm P}$($T$) curve weaken. However, both phase transitions persist up to $9$ T (the highest applied magnetic field), indicating robust antiferromagnetic couplings in this system.

\subsection{Ferroelectricity and magnetoelectricity}
According to established knowledge of magnetoelectricity, the cycloidal magnetic structure can break the spatial inversion symmetry and lead to the magnetic ferroelectricity \cite{Katsura:Prl}. In the following, the multiferroicity of LiFe(WO$_4$)$_2$ is studied.

\begin{figure}
\centering
\includegraphics[width=0.48\textwidth]{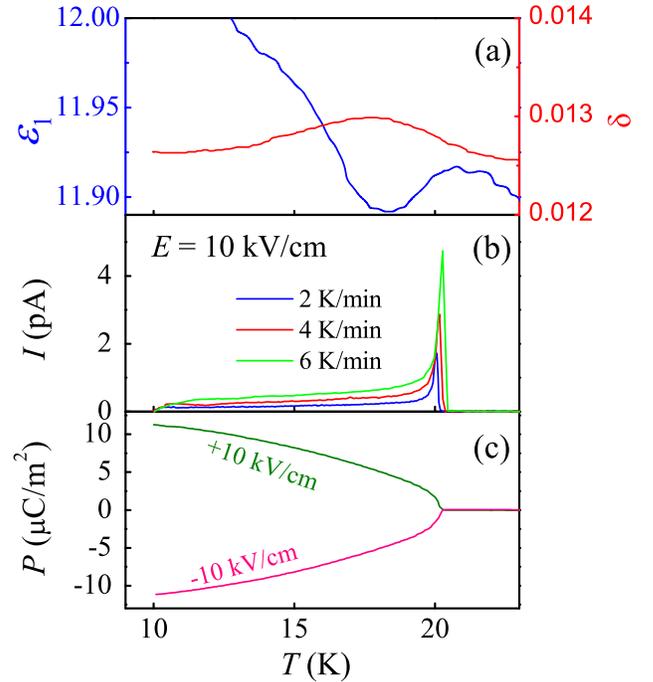}
\caption{(a) Dielectric constant (left) and dielectric loss (right). (b) Pyroelectric currents measured with different warming rates. The poling electric field is $10$ kV/cm. (c) Integrated pyroelectric $P$'s with positive/negative poling fields. The peak position of dielectric constant coincides with the emergence of pyroelectric $P$'s.}
\label{Fig5}
\end{figure}

First, the dielectric constant $\varepsilon$($T$) measured at $1$ kHz (Fig.~\ref{Fig5}(a), left axis) shows a broad peak around $T_{\rm N2}$, which is an indication of ferroelectricity. The dielectric loss (Fig.~\ref{Fig5}(a), right axis) is very small, implying negligible leakage and reliable pyroelectric measurement. Then the pyroelectric curves ($I_{pyro}$-$T$) with three warming rates ($2$, $4$, and $6$ K/min) are shown in Fig.~\ref{Fig5}(b). The three peaks of $I_{pyro}$-$T$ curves are exactly at the identical position without any shift. The $I_{pyro}$-$T$ curves are also measured under the positive and negative pooling electrical fields ($E=\pm1000$ kV/m). The integrated $P$($T$) curves are as shown in Fig.~\ref{Fig5}(c). The symmetrical $P$($T$) curves upon the positive/negative poling fields suggest the reversibility of $P$. According to $P$($T$) and $\varepsilon$($T$), the ferroelectricity emerge just below $T_{\rm N2}$, as expected from NPD. The value of $P$ at $10$ K is about $15$ $\mu$C/m$^2$ for the polycrystalline sample. Considering the value of $P$ is only $50$ $\mu$C/m$^2$ for single crystalline MnWO$_4$ \cite{Taniguchi:Prl}, the intrinsic saturated $P$ of LiFe(WO$_4$)$_2$ should be in the same order of magnitude with or even higher than that of MnWO$_4$.

\begin{figure}
\centering
\includegraphics[width=0.48\textwidth]{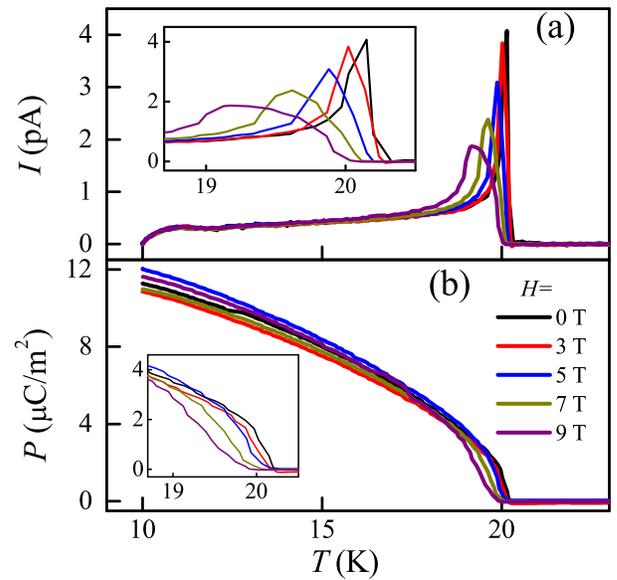}
\caption{Magnetoelectricity. (a) Pyroelectric currents measured under different magnetic fields. (b) The corresponding pyroelectric $P$'s. Inserts: magnified views around $T_{\rm N2}$.}
\label{Fig6}
\end{figure}

The coincidence of magnetic $T_{\rm N2}$ and ferroelectric $T_{\rm C}$ is a strong evidence for magnetism driven ferroelectricity, which is expected to arise from the cycloidal magnetism revealed by neutron diffraction experiments. To further confirm the intrinsic magnetoelectricity, a series of $I_{pyro}$($T$) curves at the warming rate $4$ K/min are measured under different magnetic fields (parallel to the poling electric field) up to $9$ T, as shown in Fig.~\ref{Fig6}(a). The corresponding $P$($T$) curves under magnetic fields are showed in Fig.~\ref{Fig6}(b). First, with increasing magnetic field, the current peak shifts to lower $T$ and become weaker and broader. Second, the evolution of ferroelectric $T_{\rm C}$ consists with the tendency of specific heat (Fig.~\ref{Fig4}(c)). All these characters imply the direct coupling between magnetic order and dipole order. The weak change of $P$ under $H$ is probably due to the fact that $H$ was applied along the poling electric field direction.

\subsection{DFT results}
To further understand the physics involved in LiFe(WO$_4$)$_2$, the DFT calculations have been performed with the experimental crystalline structure.

\begin{table}
\centering
\caption{Summary of DFT results of LiFeW$_2$O$_8$. Five collinear magnetic configurations (A-E) and the $\textbf{k}$=($1$, $0$, $1/3$) noncollinear (NC) order are considered. ``+" and ``-" denote spin up and down, respectively. The indices of Fe can be found in Fig.~\ref{Fig1}(b). All energies (in unit of meV/Fe) are obtained in the conditions of SOC enabled and $U_{\rm eff}=4$ eV. The lowest energy one (NC) is set as the energy reference point.}
\label{TableIII}
\begin{tabular*}{0.48\textwidth}{@{\extracolsep{\fill}}|c|c|c|c|c|c|c|c|c|c|}
\hline
\multirow{2}{*}{Magnetism} & \multicolumn{8}{c|}{Magnetic ions}                & \multirow{2}{*}{Energy}\\ \cline{2-9}
                         & Fe1 & Fe1' & Fe2 & Fe2' & Fe3 & Fe3' & Fe4 & Fe4' &                                                   \\ \hline
A                        & +   & +    & -   & -    & -   & -    & +   & +    & $10.98$                                             \\ \hline
B                        & +   & +    & -   & -    & +   & +    & -   & -    & $6.46$                                              \\ \hline
C                        & +   & +    & +   & +    & -   & -    & -   & -    & $1.89$                                            \\ \hline
D                        & +   & +    & +   & +    & +   & +    & +   & +    & $15.54$                                            \\ \hline
E                        & +   & -    & +   & -    & -   & +    & -   & +    & $1.55$                                            \\ \hline
NC                       & \multicolumn{8}{c|}{$\textbf{k}=$($1$, $0$, $1/3$)} & $0$                                            \\ \hline
\end{tabular*}
\end{table}

First, to reveal the magnetic couplings in LiFe(WO$_4$)$_2$, various magnetic orders are calculated, as summarized in Table~\ref{TableIII}. Five collinear plus one noncollinear magnetic states are considered. Although the direct adopting of experimental magnetic state is practically unavailable for the DFT calculation due to the incommensurate propagation vector, the noncollinear one adopted in our calculation, with $\textbf{k}=$($1$, $0$, $1/3$), is quite close to the experimental one. Such an approximation is acceptable according to previous experience of DFT calculations on TbMnO$_3$ and CaMn$_7$O$_{12}$ \cite{Malashevich:Prl,Xiang:Prl,Lu:Prl}. Indeed, this noncollinear state owns the lowest energy comparing with other candidates.

Using the energies of five collinear magnetic states, the exchange couplings $J$'s between Fe spins can be roughly estimated by mapping LiFeW$_2$O$_8$ to a classical Heisenberg model ($-J_{ij}S_i\cdot S_j$). Here the normalized value $\left|S\right|=1$ is used. The fitted exchange $J$'s between neighbor Fe ions (indicated in Fig.~\ref{Fig1}(b)): $J_1=2.28$ meV, $J_2=-2.44$ meV, and $J_3=-1.13$ meV. Only $J_1$ is ferromagnetic and the antiferromagnetic $J_2$ and $J_3$ frustrate the magnetism.

\begin{figure}
\centering
\includegraphics[width=0.48\textwidth]{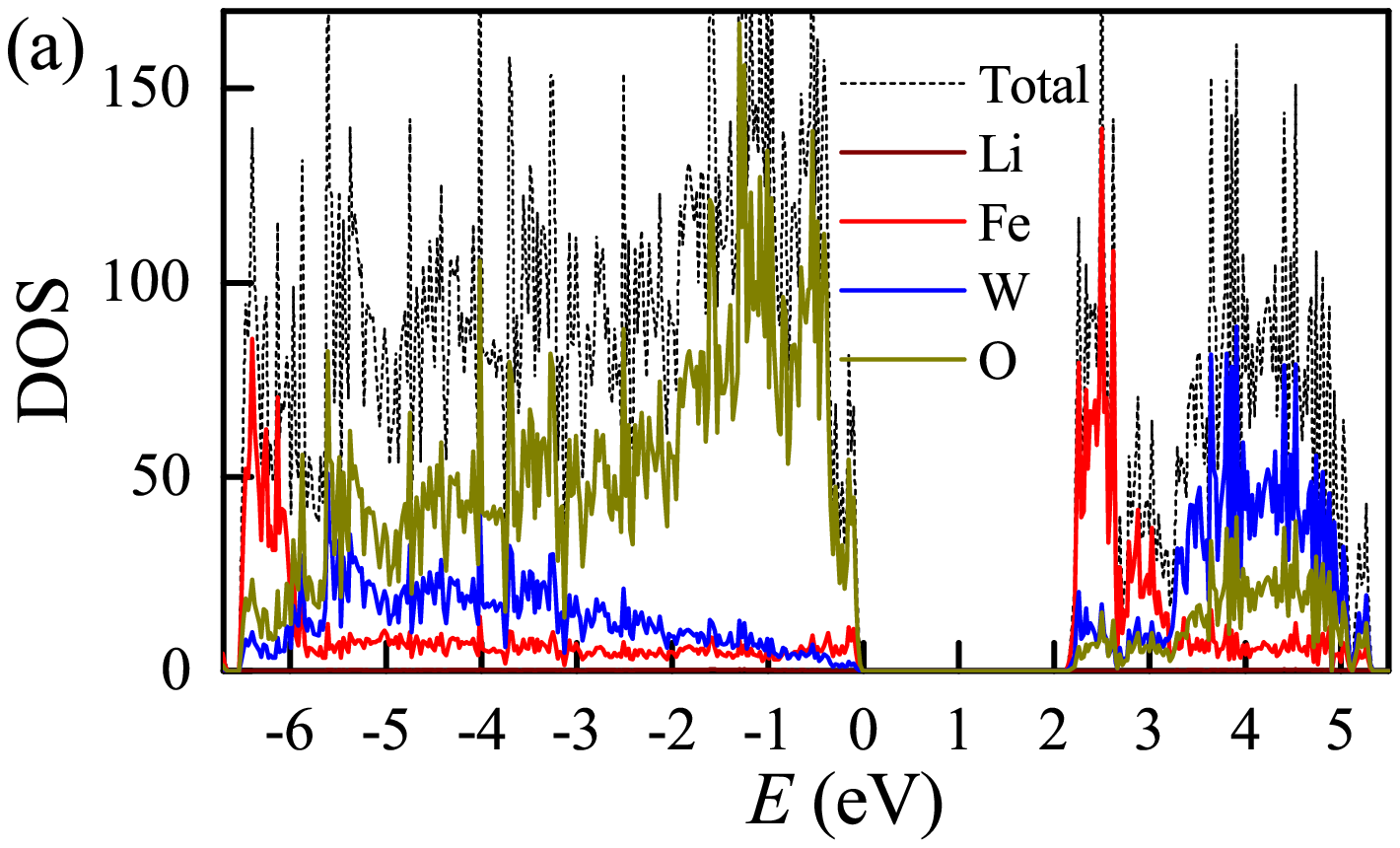}
\includegraphics[width=0.45\textwidth]{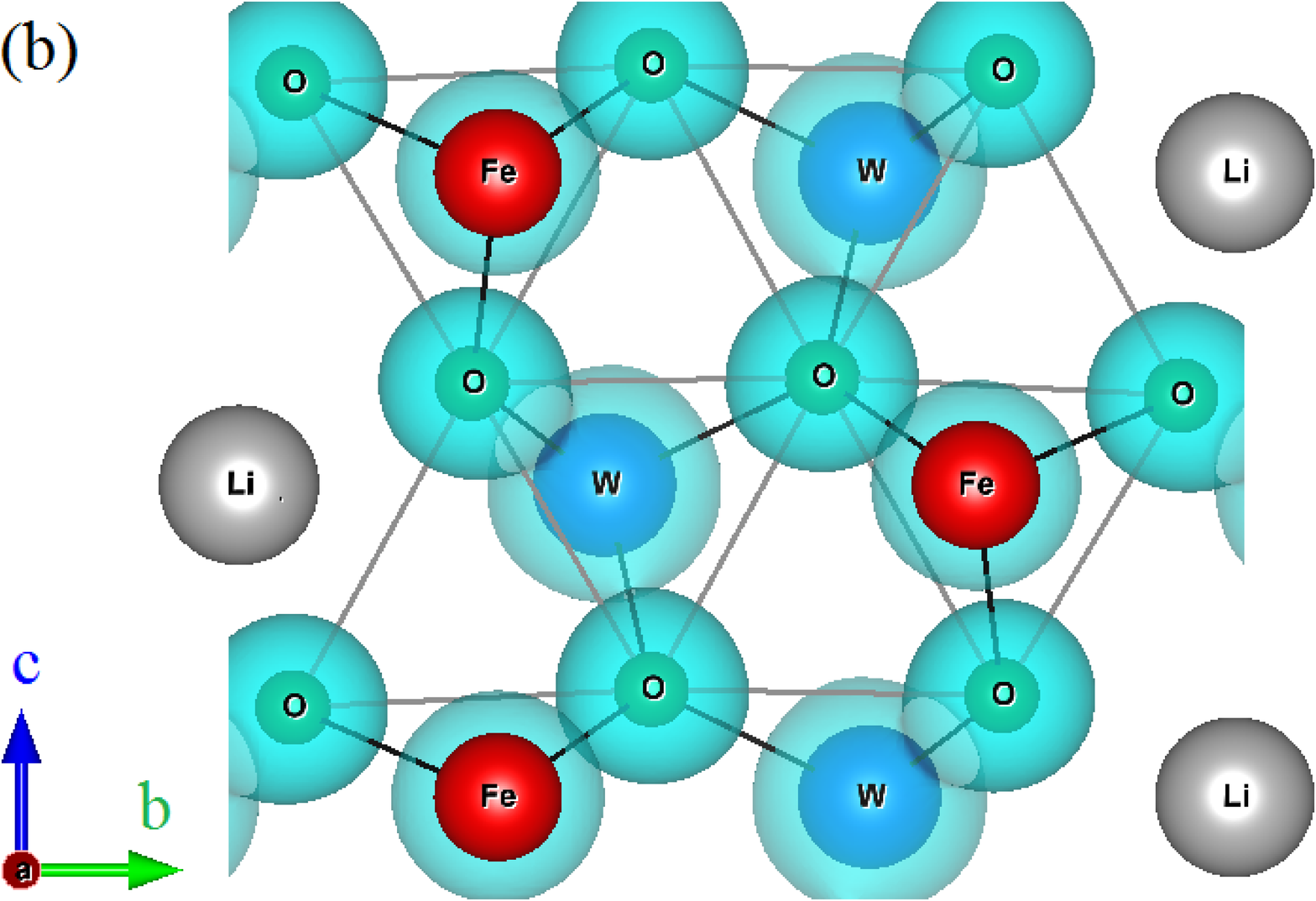}
\caption{DFT results of LiFe(WO$_4$)$_2$ with non-collinear magnetic state, $\textbf{k}=$($1$, $0$, $1/3$). (a) The density of states (DOS). (b) Contour plot of charge density. The nearest neighbor Fe-O, W-O, and O-O are connected.}
\label{Fig7}
\end{figure}

The density of states (DOS) and the spatial distribution of electronic density for the noncollinear state are shown in Fig.~\ref{Fig7}. The states near the Fermi level are mostly contributed by O's $2p$ orbitals, which hybridize with Fe's $3d$ orbitals and W's $5d$ orbitals. Although nominally the valence of W is $+6$ with empty $5d$ orbitals, our DOS diagram still suggests the involvement of W's $5d$ orbitals below the Fermi level. The band gap is about $2.4$ eV, a good insulator. As shown in Fig.~\ref{Fig7}(b), the magnetic coupling between Fe ions can be mediated via Fe-O-O-Fe and Fe-O-W-O-Fe, instead of Fe-O-Li-O-Fe.

As stated before, the cycloidal magnetic structure can lead to ferroelectric $P$, which can be confirmed by the DFT calculation too. Using the standard Berry phase method, the calculation with SOC gives a finite $P$ about $24.5$ $\mu$C/m$^2$, which is qualitatively consistent with the experimental pyroelectric value ($\sim12$ $\mu$C/m$^2$). The direction of $P$ is along the [010] axis, in consistent with the Katsura-Nagaosa-Balatsky's theory ($P\sim e_{ij}\times(S_i\times S_j)$) for noncollinear spin order \cite{Katsura:Prl}. 
This small magnitude of $P$ is also reasonable considering the nominally high-spin $3d^5$ configuration of Fe ions. The first-order SOC is quenched due to the zero orbital moment ($L=0$). Only higher-order SOC, i.e. the hybridization between Fe's $3d$ and W's $5d$ via O's $2p$, is allowed, which is thus weak. This $P$ is in the same order of magnitude comparing with that of MnWO$_4$ ($\sim50$ $\mu$C/m$^2$ for single crystalline sample) \cite{Taniguchi:Prl}, which also owns the $3d^5$ configuration.

Finally, it should be noted that the comparison of calculated $P$ and experimental $P$ is only qualitative, namely the direction and order of magnitude are reliable. The experimental $P$ can be affected by the polycrystalline grain boundaries and non-saturation. The calculated $P$ here is from pure electronic contribution, while the ion displacements has not been taken into account since the SOC-enabled lattice optimization is too CPU-demanding for such a large cell. So both the experimental $P$ and calculated $P$ are not the precise value of intrinsic saturated $P$. Even though, the qualitative comparison remains valuable and consistent.

\subsection{Additional discussion}
Above experimental measurements were done with polycrystalline samples. Usually, single crystals are better to characterize the multiferroicity, for both the neutron scattering and electric measurements. In this sense, further studies based on single crystals are highly encouraged. Even though, the current study have shown strong evidences and consistent results regarding the multiferroicity from several aspects.

\section{Conclusion}
The physical properties of LiFe(WO$_4$)$_2$, especially the magnetism and multiferroicity, have been systematically investigated. Sequential magnetic transitions at $T_{\rm N1}=21.6$ K and $T_{\rm N2}=19.7$ K are observed. Below $T_{\rm N2}$, a ferroelectric polarization emerges in LiFe(WO$_4$)$_2$ which is driven by particular noncollinear magnetism and thus can be tuned by magnetic field. Such type-II multiferroicity is verified by both neutron diffraction and density functional theory. According to our study, LiFe(WO$_4$)$_2$ is the second confirmed multiferroic member in the $A$WO$_4$ tungstates family, different from the non-multiferroic NaFe(WO$_4$)$_2$. Comparing with the first confirmed multiferroic MnWO$_4$, LiFe(WO$_4$)$_2$ owns higher and wider ferroelectric temperature, while its polarization is in the same order of magnitude. Our study will hopefully stimulate more studies on the tungstates family.

\acknowledgments{Work was supported by the National Key Research Programs of China (Grant No. 2016YFA0300101) and National Natural Science Foundation of China (Grant Nos. 11674055, 11374147, 11374112). V. O. Garlea and T. Zou (ORNL) acknowledge the support from the Scientific User Facilities Division, Office of Basic Energy Sciences, US Department of Energy (DOE). Most calculations were done on Tianhe-2 at National Supercomputer Centre in Guangzhou (NSCC-GZ).

\bibliographystyle{apsrev4-1}
\bibliography{../ref3}
\end{document}